\documentclass[prl,twocolumn,noshowpacs,nobalancelastpage,tightenlines,floatfix]{revtex4}
\usepackage{amssymb}
\usepackage{graphicx}
\begin{document}

\title{Cavity QED ``By The Numbers''}
\author{J.~McKeever, J.~R.~Buck, A.~D.~Boozer, and H.~J.~Kimble}
\affiliation{Norman Bridge Laboratory of Physics 12-33\\
California Institute of Technology, Pasadena, CA 91125}
\date{\today}

\begin{abstract}
The number of atoms trapped within the mode of an optical cavity is
determined in real time by monitoring the transmission of a weak probe beam.
Continuous observation of atom number is accomplished in the strong coupling
regime of cavity quantum electrodynamics and functions in concert with a
cooling scheme for radial atomic motion. The probe transmission exhibits
sudden steps from one plateau to the next in response to the time evolution
of the intracavity atom number, from $N\geq 3$ to $N=2 \rightarrow 1
\rightarrow 0$, with some trapping events lasting over $1$ second.
\end{abstract}

\maketitle

Cavity quantum electrodynamics (QED) provides a setting in which atoms
interact predominantly with light in a single mode of an electromagnetic
resonator of high quality factor $Q$ \cite{kimble98}. Not only can the light
from this mode be collected with high efficiency \cite{spg04}, but as well
the associated rate of optical information for determining atomic position
can greatly exceed the rate of free-space fluorescent decay employed for
conventional imaging \cite{hood00}. Moreover the regime of strong coupling,
in which coherent quantum interactions between atoms and cavity field
dominate dissipation, offers a unique setting for the study of open quantum
systems \cite{mabuchi02}. Dynamical processes enabled by strong coupling in
cavity QED provide powerful tools in the emerging field of quantum
information science (QIS), including for the implementation of quantum
computation \cite{pellizzari95} and for the realization of distributed
quantum networks \cite{cirac97,briegel00}.

With these prospects in mind, experiments in cavity QED have made
made great strides in trapping single atoms in the regime of
strong coupling \cite{ye99,hood00,trapping03,maunz04}. However,
many protocols in QIS require multiple atoms to be trapped within
the same cavity, with \textquotedblleft quantum
wiring\textquotedblright\ between internal states of the various
atoms accomplished by way of strong coupling to the cavity field
\cite{pellizzari95,duan03,hong02,sorensen03}. Clearly the
experimental ability to determine the number of trapped atoms
coupled to a cavity is a critical first step toward the
realization of diverse goals in QIS. Experimental efforts to
combine ion trap technology with cavity QED are promising
\cite{ions}, but have not yet reached the regime of strong
coupling.

\begin{figure}[tb]
\includegraphics[width=8.6cm]{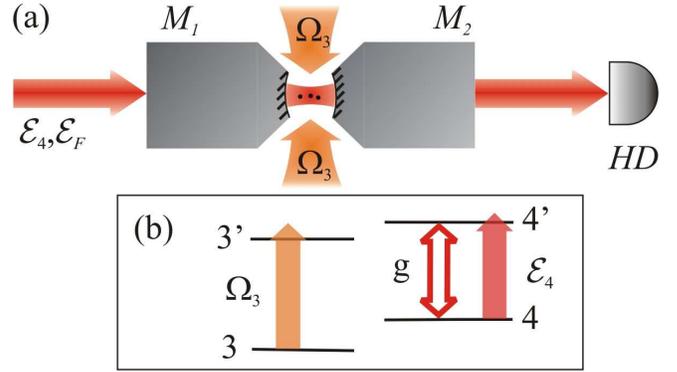}
\caption{(color online) Schematic of our experiment. Cs atoms are loaded
into an intracavity FORT ($\mathcal{E}_{F}$) by way of the transverse
cooling field $\Omega _{3}$ and the cavity probe field $\mathcal{E }_{4}$.
The transmitted $\mathcal{E}_{4}$ field is directed to a heterodyne detector
(\textit{HD}), allowing real-time determination of intracavity atom number.}
\label{setup}
\end{figure}

In this Letter we report measurements in which the number of atoms
trapped inside an optical cavity is observed in real time. After
initial loading of the intracavity dipole trap with
$\bar{N}\approx 5$ atoms, the decay of atom number $N\geq
3\rightarrow 2\rightarrow 1\rightarrow 0$ is monitored by way of
changes in the transmission of a near-resonant probe beam, with
the transmitted light exhibiting a cascade of \textquotedblleft
stairsteps\textquotedblright\ as successive atoms leave the trap.
After the probabilistic loading stage, the time required for the
determination of a particular atom number $N=1,2,3$ is much
shorter than the mean interval over which the $N$ atoms are
trapped. Hence, a precise number of intracavity atoms can be
prepared for experiments in QIS, for which the timescales
$(g^{-1}\approx 10^{-8}~\mathrm{s})\ll (\tau \approx 3~
\mathrm{s}$), where $\tau$ is the atomic trapping time
\cite{trapping03} and $\hbar g$ is the atom-field interaction
energy. In the present case, the atom number is restricted to
$N\lesssim 3$, but the novel detection scheme that we have
developed may enable extensions to moderately larger atom numbers
$N\lesssim 10$.

As illustrated in Fig. \ref{setup}, our experiment combines laser
cooling, state-insensitive trapping, and strong coupling in cavity
QED, as were initially achieved in Ref. \cite{trapping03}. A cloud
of Cs atoms is released from a magneto-optical trap (MOT) several
$\mathrm{mm}$ above the cavity, which is formed by the reflective
surfaces of mirrors $(M_{1},M_{2})$ . Several atoms are cooled and
loaded into an intracavity far-off-resonance trap (FORT) and are
thereby strongly coupled to a single mode of the cavity. The
maximum single-photon Rabi frequency $2g_{0}$ for one atom is
given by $ g_{0}/2\pi =24~\mathrm{MHz}$, and is based on the
reduced dipole moment for the $6S_{1/2},F=4\rightarrow
6P_{3/2},F^{\prime }=4^{\prime }$ transition of the $D2$ line in
Cs at $\lambda_{0}=852.4~\mathrm{nm}$. Decay rates for the
$6P_{3/2}$ atomic excited states and the cavity mode at
$\omega_{0}=2\pi c/\lambda_{0}$ are $\gamma
/2\pi=2.6~\mathrm{MHz}$ and $\kappa /2\pi=4.2~ \mathrm{MHz}$,
respectively. The fact that $g_{0}\gg (\kappa,\gamma)$ places our
system in the strong coupling regime of cavity QED
\cite{kimble98}, giving critical atom and photon numbers
$n_{0}\equiv \gamma ^{2}/(2g_{0}^{2})\approx 0.0057$, $N_{0}\equiv
2\kappa \gamma /g_{0}^{2}\approx 0.037$.

The cavity is independently stabilized and tuned such that it
supports $ TEM_{00}$ modes simultaneously resonant with both the
$F=4\rightarrow F^{\prime }=4^{\prime }$ atomic transition at
$\lambda _{0}$ and our FORT laser at $\lambda
_{F}=935.6~\mathrm{nm}$, giving a length $l_{0}=42.2~ \mathrm{\mu
m}$. A weak probe laser $\mathcal{E}_{4}$ excites the cavity mode
at $\lambda _{0}$ with the cavity output directed to detector
\textit{HD}, while a much stronger trapping laser
$\mathcal{E}_{F}$ drives the mode at $\lambda _{F}$. In addition,
the region between the cavity mirrors is illuminated by two
orthogonal pairs of counter-propagating cooling beams in the
transverse plane (denoted $\Omega _{3}$). Atoms arriving in the
region of the cavity mode are exposed to the
$(\mathcal{E}_{4},\mathcal{E} _{F},\Omega _{3})$ fields
continuously, with a fraction of the atoms cooled and loaded into
the FORT by the combined actions of the $\mathcal{E}_{4}$ and
$\Omega _{3}$ fields \cite{trapping03}. For all measurements, the
cavity detuning\ from the $4\rightarrow 4^{\prime}$ atomic
resonance is $\Delta _{C}=0$. The detuning of the
$\mathcal{E}_{4}$ probe from the atom-cavity resonance is $\Delta
_{4}=+4~\mathrm{MHz}$, and its intensity is set such that the mean
intracavity photon number $\bar{n}=0.02$ with no atoms in the
cavity. The detuning of the $\Omega _{3}$ transverse cooling field
is $ \Delta _{3}=+25~\mathrm{MHz}$ from the $F=3\rightarrow
F^{\prime }=3^{\prime }$ resonance, and its intensity is about
$I_{3}\approx 4\times 10^{1}~ \mathrm{mW/cm^{2}}$.

The field $\mathcal{E}_{F}$\ that drives the standing-wave,
intracavity FORT is linearly polarized, resulting in nearly equal
ac-Stark shifts for all Zeeman sublevels of the $F=3,4$ hyperfine
ground states of the $6S_{1/2}$ manifold \cite{corwin99}. The peak
value of the trapping potential is $ -U_{0}/h=-47~\mathrm{MHz}$,
giving a trap depth $U_{0}/k_{B}=2.2~\mathrm{mK}$ . A critical
characteristic of the FORT is that all states within the
$6P_{3/2}$ excited manifold likewise experience a
\textit{trapping} shift of roughly $-U_{0}$ (to within $\approx
\pm 15\%$) \cite{katori99,ido00,kimble99,trapping03}, which
enables continuous monitoring of trapped atoms in our cavity and
avoids certain heating effects.

\begin{figure}[tb]
\includegraphics[width=8.6cm]{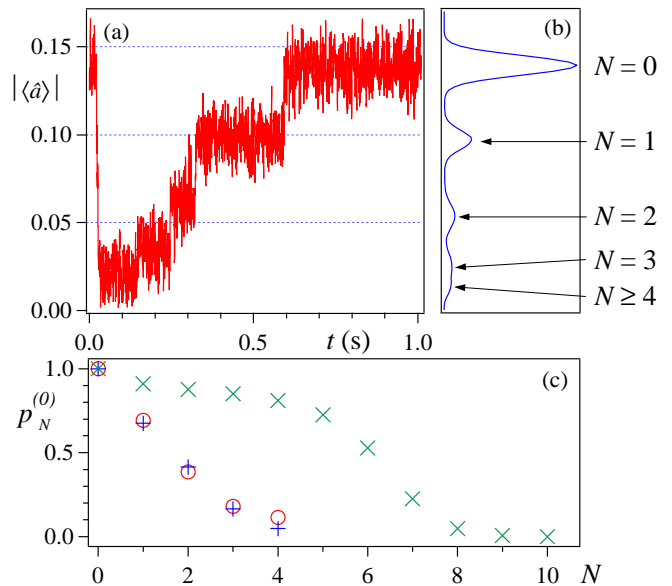}
\caption{(color online) (a) A typical detection record in which
several ($N>4 $) atoms are loaded into the trap. Heterodyne
detection bandwidth is $1~ \mathrm{kHz}$. (b) Histogram of 500
such traces, binned with respect to the heterodyne signal
$|\langle \hat{a}\rangle |$. A digital low-pass filter of
bandwidth $100~\mathrm{Hz}$ is applied to each trace prior to the
computation of the histogram. (c) Comparison of the model
prediction for $p_{0}^{(N)}(y=0.5)$ ($+$) with the measured
positions of the histogram peaks in (b) ($\bigcirc$). Also shown
($\times$) is $p_{0}^{(N)}(y=0.1)$ to indicate the possibility to
detect specific atom numbers for larger $\bar{N} $.}
\label{tr+hist}
\end{figure}

Figure \ref{tr+hist}(a) displays a typical record of the
heterodyne current $i(t)$ resulting from one instance of FORT
loading. Here, the current $i(t)$ is referenced to the amplitude
of the intracavity field $|\langle \hat{a} \rangle |$ by way of
the known propagation and detection efficiencies. The initial
sharp drop in $|\langle \hat{a}(t)\rangle |$ around $t=0$ results
from atoms that are cooled and loaded into the FORT by the
combined action of the $(\mathcal{E}_{4},\Omega _{3})$ fields
\cite{trapping03}. Falling atoms are not exposed to
$\mathcal{E}_{4}$ until they reach the cavity mode, presumably
leading to efficient trap loading for atoms that arrive at a
region of overlap between the standing waves at
$(\lambda_{0},\lambda_{F})$ for the
$(\mathcal{E}_{4},\mathcal{E}_{F})$ fields. Trap loading always
occurs within a $\pm 10~\mathrm{ms}$ window around
$t=0.025~\mathrm{s}$ (relative to $t=0$ in Fig. \ref{tr+hist}(a)).
This interval is determined using separate measurements of the
arrival time distribution of freely falling atoms in the absence
of the FORT \cite{mabuchi96,ye99}.

Subsequent to this loading phase, a number of remarkable features are
apparent in the trace of Fig. \ref{tr+hist}(a), and are consistently present
in all the data. The most notable characteristic is the fact that the
transmission versus time consists of a series of flat ``plateaus'' in which
the field amplitude is stable on long timescales. Additionally, these
plateaus reappear at nearly the same heights in all repeated trials of the
experiment, as is clearly evidenced by the histogram of Fig. \ref{tr+hist}
(b). We hypothesize that each of these plateaus represents a different
number $N$ of trapped atoms coupled to the cavity mode, as indicated by the
arrows in Fig. \ref{tr+hist}.

Consider first the one-atom case, which unexpectedly exhibits
relatively large transmission and small variance. For fixed drive
$\mathcal{E}_{4}$, the intracavity field is a function of the
coupling parameter $g^{(i,f)}(\mathbf{\ r})=g_{0}G_{i,f}\sin
(k_{0}z)\exp (-2\rho ^{2}/w_{0}^{2})$ where $\rho $ is the
transverse distance from the cavity axis ($z$), $k_{0}=2\pi
/\lambda _{0}$, and $G_{i,f}$ relates to the Clebsch-Gordan
coefficient for particular initial and final states $(i,f)$ within
the $F=4,F^{\prime }=4^{\prime }$ manifolds \cite{hood98,puppe03}.
Variations in $g$ as a function of the atom's position
$\mathbf{r}$ and internal state might reasonably be expected to
lead to large variations of the intracavity field, both as a
function of time and from atom to atom.

However, one atom in the cavity produces a reasonably well-defined
intracavity intensity $I\propto |\langle \hat{a}\rangle |^{2}$ due
to the interplay of two effects. The first is that for small probe
detunings $\Delta _{4}$, the intracavity intensity $I_{1}$ for one
atom is suppressed by a factor $f$ relative to the empty-cavity
intensity $I_{0}$, where for weak excitation, $f\approx
4C_{1}^{2}\gg 1$ with $C_{1}=g^{2}/2\kappa \gamma $. A persistent,
strongly reduced transmission thereby results, since the condition
$[C_{1}^{(i,f)}(\mathbf{r})]^{2}\gg 1$ is robust to large
fluctuations in atomic position $\mathbf{r}$ and internal state.
The second effect is that the $F=4\leftrightarrow F^{\prime
}=4^{\prime }$ transition cannot be approximated by a closed,
two-level system, since the $F^{\prime }=4^{\prime }$ excited
states decay to both $F=3,4$ hyperfine ground levels. As
illustrated in Fig. \ref{setup}(b), an atom thus spends a fraction
$q$ of its time in the cavity QED manifold $(4,4^{\prime })$, and
a fraction $p\approx 1-q$ in the $(3,3^{\prime })$ manifold. In
this latter case, the effective coupling is negligible
($C_{1}^{eff}\approx 4\times 10^{-4})$, leading to an intensity
that approximates $I_{0}$. Hence, the intracavity intensity as a
function of time $I(t)$ should have the character of a random
telegraph signal switching between levels $(I_{0},I_{1})$, with
dwell times determined by $(\mathcal{E}_{4},\Omega _{3})$, which
in turn set $p$ \cite {dark}. Since $(\mathcal{E}_{4},\Omega
_{3})$ drive their respective transitions near saturation, the
timescale $\tau _{P}\sim 1~\mu \mathrm{s}$ for optical pumping
from one manifold to another is much faster than the inverse
detection bandwidth $(1/2\pi B)\approx 160~\mu \mathrm{s}$. The
fast modulation of the intracavity intensity due to optical
pumping processes thereby gives rise to an average detected signal
corresponding to intensity $\bar{I}_{1}\approx
pI_{0}+qI_{1}\approx pI_{0}$ for $I_{1}\ll I_{0}$.

This explanation for the case of $1$ atom can be extended to $N$
intracavity atoms to provide a simple model for the
\textquotedblleft stairsteps\textquotedblright\ evidenced in Fig.
\ref{tr+hist}(a). For $N$ atoms, the intracavity intensity should
again take the form of a random telegraph signal, now switching
between the levels $(I_{0},I_{k})$, with high transmission $I_{0}$
during intervals when all $N$ atoms happen to be pumped into the
$(3,3^{\prime })$ manifold, and with low transmission $I_{k}\leq
I_{1}$ anytime that $1\leq k\leq N$ atoms reside in the
$(4,4^{\prime })$ manifold, where $I_{k}\sim I_{1}/k^{2}$ for weak
excitation with $\Delta _{C}=\Delta _{4}=0$. The intracavity
intensities $\{I_{k}\}$ determine the transition rates $\{\gamma
_{k\rightarrow k-1}\}$ between states with $k$ and $k-1$ atoms in
the $(4,4^{\prime })$ manifold, while $\Omega _{3}$ determines
$\{\gamma _{k-1\rightarrow k}\}$ for $k-1\rightarrow $ $k$ via
transitions from the $(3,3^{\prime })$ manifold. For the hierarchy
of states $k=0,1,\ldots ,N$ with transition rates $\{\gamma
_{k\rightarrow k-1,}\gamma _{k-1\rightarrow k}\}$, it is
straightforward to determine the steady-state populations
$p_{k}^{(N)}$. With the physically motivated assignments $\gamma
_{k-1\rightarrow k}=\gamma _{0\rightarrow 1}$ independent of $k$
and $\gamma _{k\rightarrow k-1}=\gamma _{1\rightarrow 0}/k^{2}$
corresponding to $I_{k}\sim I_{1}/k^{2}$, we find that
$p_{0}^{(N)}=1/\sum_{k=0}^{N}(k!)^{2}y^{k}$, where $y\equiv \gamma
_{0\rightarrow 1}/\gamma _{1\rightarrow 0}$. Hence, for $I_{k}\ll
I_{0}$, the prediction for the average intensity is
$\bar{I}_{N}\approx p_{0}^{(N)}I_{0}$, which leads to a sequence
of plateaus of increasing heights $\bar{I}_{N+1}\rightarrow
\bar{I}_{N}\rightarrow \bar{I}_{N-1}$ as successive atoms are lost
from the trap $N+1\rightarrow N\rightarrow N-1$.

Figure \ref{tr+hist}(c) compares the prediction\ of this simple
model with the measured values of peak positions in (b). The only
adjustable parameter is the value $y=0.5$, resulting in reasonable
correspondence between the model and the measurements. Also shown
are values $p_{0}^{(N)}$ for $y=0.1$ to indicate that it might be
possible to enhance the resolution for a particular range of atom
number by framing a given few values $N_{1},N_{1}\pm 1$ in the
transition region $p_{0}^{(N_{1})}\approx 0.5$, where
$N_{1}\approx 6$ in (c). This could be accomplished by adjusting
the relative strengths of the $(\mathcal{E}_{4},\Omega _{3})$
fields and hence $y $.

Although our simple model accounts for the qualitative features in
Fig. \ref{tr+hist}, a quantitative description requires a
considerably more complex analysis based upon the full master
equation for $N$ intracavity atoms, including the multiple Zeeman
states and atomic motion through the polarization gradients of the
$\Omega _{3}$ beams. We have made initial efforts in this
direction \cite{boozer04} for one atom, and are working to extend
the treatment to $N\geq 2$ atoms.

\begin{figure}[tb]
\includegraphics[width=8.6cm]{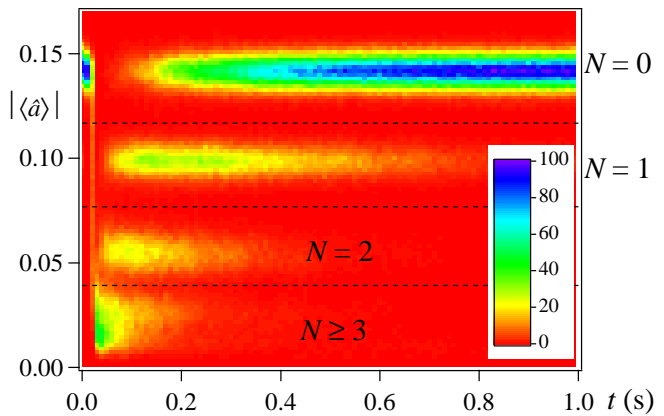}
\caption{(color online) Histogram of 500 traces such as the one in
Fig. \protect\ref{tr+hist}(a), binned with respect to both signal
strength $|\langle \hat{a}\rangle|$ and time $t$. The signals are
filtered first as in Fig. \protect\ref{tr+hist}(b).}
\label{2dhist}
\end{figure}

Beyond these considerations, additional evidence that the plateaus
in Fig. \ref{tr+hist} correspond to definite atom numbers is
provided by Fig. \ref{2dhist}. Here, the data recorded for the
probe transmission have been binned not only with respect to the
value of $|\langle \hat{a}\rangle |$ as in Fig. \ref{tr+hist}(b),
but also as a function of time. Definite plateaus for $|\langle
\hat{a}\rangle |$ are again apparent, but now their characteristic
time evolution can be determined. The critical feature of this
plot is that the plateaus lying at higher values of $|\langle
\hat{a} \rangle |$ correspond to times \textit{later} in the
trapping interval, in agreement with the expectation that $N$
should always decrease with time beyond the small window of trap
loading around $t=0.025~\mathrm{s}$. This average characteristic
of the entire data set supports our hypothesis that the plateaus
in $|\langle \hat{a}\rangle |$ correspond to definite intracavity
atom numbers $N$, as indicated in Figs. \ref{tr+hist} and
\ref{2dhist}. Moreover, none of the 500 traces in the data set
includes a downward step in transmission after the initial trap
loading.

\begin{figure}[tb]
\includegraphics[width=8.6cm]{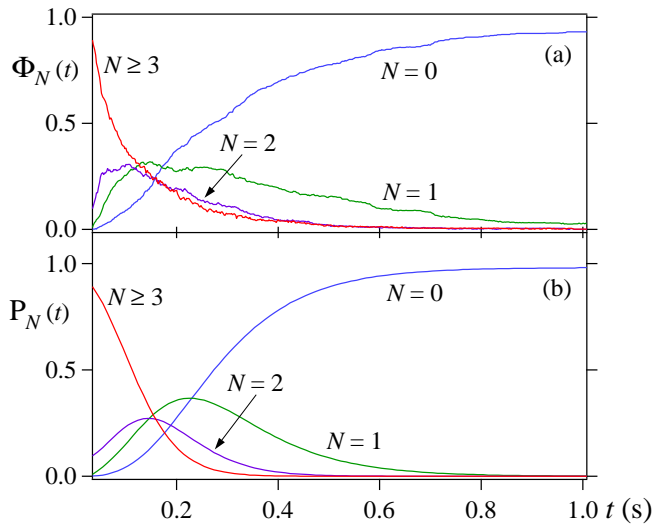}
\caption{(color online) (a) Experimental results for the time evolution of
the $N$-atom populations $\Phi _{N}(t)$, which are normalized such that
their sum is approximately unity throughout the interval shown. (b) The
results of a simple model calculation $P_{N}(t)$ are fit to the data $\Phi
_{N}(t)$ with one free parameter, the single atom decay rate $\Gamma $.}
\label{bumpdecay}
\end{figure}

To examine the dynamics of the trap loss more quantitatively, we
consider each atom number individually by integrating the
\textquotedblleft plateau\textquotedblright\ regions along the
$|\langle \hat{a}(t)\rangle |$ axis for each time $t$. The dashed
horizontal lines in Fig. \ref{2dhist} indicate the boundaries
chosen to define the limits of integration for each value of $N$.
We thereby obtain time-dependent ``populations'' $\Phi _{N}(t)$
for $N=0,1,2$, and $\Phi _{\geq 3}(t)=\sum_{N=3}^{\infty }\Phi
_{N}(t)$, which are plotted in Fig. \ref{bumpdecay}(a). To isolate
the decay dynamics from those of trap loading, we plot the data
beginning at $t_{0}=0.034~ \mathrm{s}$ with respect to the origin
in Figs. \ref{tr+hist}(a) and \ref{2dhist}. The qualitative
behavior of these populations is sensible, since almost all trials
begin with $N\geq 3$, eventually decaying to $N=2,1,0$.

The quantities $\Phi _{N}(t)$ are approximately proportional to
the fraction of experimental trials in which $N$ atoms were
trapped at time $t$, so long as the characteristic duration
$\Delta t_{N}$ of each plateau far exceeds the time resolution of
the detection. If the bandwidth is too low, transient steps no
longer represent a negligible fraction of the data, as is the case
for transitions between the shortest-lived levels (e.g.,
$N=3\rightarrow 2$). We estimate that this ambiguity causes
uncertainties in $\Phi _{N}$ at the $5-10\%$ level.

Also shown in panel (b) of Fig. \ref{bumpdecay} is the result of a
simple birth-death model for predicting the time evolution of the
populations, namely $\dot{P}_{N}(t)=-\Gamma
(NP_{N}(t)-(N+1)P_{N+1}(t))$, where $P_{N}(t)$ represents the
probability of $N$ atoms in the trap. The main assumption of the
model is that there is one characteristic decay rate $\Gamma $ for
trapped atoms, and that each atom leaves the trap independently of
all others. Initial conditions for $N=0$, $1$ and $2$ for the
solution presented in Fig. \ref{bumpdecay}(b) are obtained
directly from the experimental data after trap loading, $\Phi
_{N}(t_{0})$. Since the plateaus for higher values of $N$ are not
well resolved, we use a Poisson distribution for $N\geq 3$. The
mean $\mu=5.2$ is obtained by solving
$\sum_{N=3}^{\infty}e^{-\mu}\mu^N/N!=\Phi _{N\geq 3}(t_{0})$.
Given these initial conditions, we perform a least-squares fit of
the set of analytic solutions $\{P_{N}(t)\} $ to the set of
experimental curves $\{\Phi _{N}(t)\}$ with $\Gamma $ the only
free parameter, resulting in the curves in Fig. \ref{bumpdecay}(b)
with $\Gamma =8.5~\mathrm{s}^{-1}$. Although there is reasonable
correspondence between Figs. \ref{bumpdecay} (a) and (b), $\Phi
_{N}(t)$ evolves more rapidly than does $P_{N}(t)$ at early times,
and yet the data decay more slowly at long times. This suggests
that there might be more than one timescale involved, possibly due
to an inhomogeneity of decay rates from atom to atom or to a
dependence of the decay rate on $N$. We have observed
non-exponential decay behavior in other measurements of
single-atom trap lifetimes, and are working to understand the
underlying trap dynamics.

Our experiment represents a new method for the real-time
determination of the number of atoms trapped and strongly coupled
to an optical cavity. We emphasize that an exact number $N=1$ to
$3$ coupled atoms can be prepared in our cavity within $\approx
200~\mathrm{ms}$ from the release of the MOT. Although the trap
loading is not deterministic, $N$ can be measured quickly compared
to the subsequent trapping time $\tau \approx 3~\mathrm{s}$ \cite
{trapping03}. These new capabilities are important for the
realization of various protocols in quantum information science,
including probabilistic protocols for entangling multiple atoms in
a cavity \cite {duan03,hong02,sorensen03}. Although our current
investigation has centered on the case of small $N\leq 3$, there
are reasonable prospects to extend our technique to higher values
$N\lesssim 10$ as, for example, by way of the strategy illustrated
in Fig. \ref{tr+hist}(c). Moreover, the rate at which we acquire
information about $N$ can be substantially increased from the
current value $\kappa |\langle \hat{a} \rangle |^{2}\sim 10^{5}/$s
toward the maximum rate for optical information $ g^{2}/\kappa
\gtrsim 10^{8}/$s, which can be much greater than the rate for
fluorescent imaging set by $\gamma$ \cite{hood00}.


\begin{thebibliography}{99}
\bibitem{dark} Fluctuations in intracavity intensity can also arise from
optical pumping into dark states. Although the $\mathcal{E}_{4}$
probe field is linearly polarized with dark state $(F=4,m_{F}=0)$,
independent measurements indicate the presence of significant
Zeeman splittings $\Delta \nu_{\pm}\approx 0.5$ MHz for $\Delta
m_{F}=\pm 1$, partially arising from residual ellipticity of the
FORT field \cite{trapping03,corwin99} and from uncompensated stray
magnetic fields. Moreover, an atom moving at $v=10~ \mathrm{cm/s}$
travels through the polarization gradients of the $\Omega_3$ field
in $\approx 4~\mu \mathrm{s}$, ensuring rapid pumping from dark
states in the $F=3$ ground level. Occupation of dark states for
extended periods is thereby precluded, as is supported by detailed
Monte-Carlo simulations for $N=1$\cite{boozer04}.
\end{thebibliography}
\end{document}